       \documentclass[twocolumn,amsmath,amssymb,10pt,superscriptaddress,a4paper,letterpaper,fleqn]{revtex4-1}
       \usepackage{amssymb}
       \usepackage{epsfig}
       \usepackage{graphicx}
       \usepackage{dcolumn}
       \usepackage{array}
       \usepackage{bm}
       \usepackage{fancyheadings}
       \usepackage{longtable}
       \usepackage{multirow}
       \usepackage{float}
       \usepackage{afterpage}
       \usepackage{color}
       
       \usepackage{amssymb}
       \usepackage{setspace}
       \usepackage{url}
       \usepackage{subfigure}
       \usepackage{booktabs}
       \usepackage{multirow}
       \usepackage{soul}
       
       \setlongtables
       \usepackage[breaklinks=true,linkbordercolor={1 1 1}]{hyperref}
       
\begin{document}
       \setcounter{page}{1}
       
       %%% **********************************************************************
       
       \title{
       %% Please do not remove the line below
       \qquad \\ \qquad \\ \qquad \\  \qquad \\  \qquad \\ \qquad \\
       %% Change title, authors, afiliation and type  your abstract
       Conservation of Isospin in Neutron-Rich Fission Fragments}
       
       \author{Ashok Kumar Jain}
       \email[Corresponding author: ]{ashkumarjain@yahoo.com}
       \affiliation{Department of Physics, Indian Institute of Technology, Roorkee-247667, India}
       
       \author{Deepika Choudhury}
       %\email[Corresponding author: ]{deepika.chry@gmail.com}
       \affiliation{Department of Physics, Indian Institute of Technology, Roorkee-247667, India}
       \affiliation{Department of Nuclear and Atomic Physics, Tata Institute of Fundamental Research, Mumbai-400005, India}
       
       \author{Bhoomika Maheshwari} 
       \affiliation{Department of Physics, Indian Institute of Technology, Roorkee-247667, India} 
       
       %\date{\today} 
       %\received{8 March 2013; revised received XX June 2013; accepted XX September 2013}
       
       \begin{abstract}
       {On the occasion of the $75^{th}$ anniversary of the fission phenomenon, we present a surprisingly simple result which highlights the important role of isospin and its conservation in neutron rich fission fragments. We have analysed the fission fragment mass distribution from two recent heavy-ion reactions $^{238}$U($^{18}$O,f) and $^{208}$Pb($^{18}$O,f) as well as a thermal neutron fission reaction $^{245}$Cm(n$^{th}$,f). We find that the conservation of the total isospin explains the overall trend in the observed relative yields of fragment masses in each fission pair partition. The isospin values involved are very large making the effect dramatic. The findings open the way for more precise calculations of fission fragment distributions in heavy nuclei and may have far reaching consequences for the drip line nuclei, HI fusion reactions, and calculation of decay heat in the fission phenomenon.}
       \end{abstract}
       
       \maketitle
       
       %%% DO NOT EDIT the following section enclosed by *****
       %%% ***************************************************
       \lhead{ND 2013 Article $\dots$}
       \chead{NUCLEAR DATA SHEETS}
       \rhead{A. Author1 \textit{et al.}}
       \lfoot{}
       \rfoot{}
       \renewcommand{\footrulewidth}{0.4pt}
       %%% ***************************************************
       
       %%% EDIT: the body of your text starts here, you may use as many \section, \subsection, \subsubsection
       %%% \begin{figure}, \begin{tabular} and \begin{equations} as needed. Please note that each \begin{}
       %%% must be closed with the corresponding \end{} and that section titles should be in capital
       %%% letters. Current text should be eventually deleted.

       \section{ INTRODUCTION}
       
       %\begin{multicols}{2}{
       
       According to Wigner, isospin, a fundamental entity in nuclear physics, can enable us to obtain the value of a physical quantity 
	which is more difficult to measure, from a quantity which is easier to measure or which has already been measured~\cite{Wigner1957}.
	 In 1960s, important work of Andersson {\it et al.,}~\cite{Anderson1962} and Fox {\it et al.,}~\cite{Fox1964} in the $A=50$ and $90$ 
	mass regions, respectively, lead to the rebirth of this quantum number~\cite{Robson1973}. 
       Lane and Soper~\cite{LaneSoper1962} extended the work of MacDonald~\cite{MacDonald1955} for light nuclei and investigated the degree of 
	isospin purity in heavier nuclei. This theoretical work suggested that the isospin may become a good quantum number in heavy nuclei. 
	The large number of excess neutrons in neutron-rich nuclei together have an absolutely pure isospin, which strongly dilutes the isospin 
	impurity of the remaining part of the system with N=Z~\cite{Robson1973,LaneSoper1962}. The fission fragments of heavy nuclei are highly neutron rich with the value of T$_3$ = (N-Z)/2 becoming very large~\cite{Robson1973} and are expected to carry a pure isospin.
       
       In this paper, we show that the conservation of isospin explains in a strikingly precise manner the observed relative yields of pairs of
	 fission fragments of each of the ($Z_1,Z_2$) partitions measured in two recent heavy ion induced fusion-fission experiments~\cite{Danu,Bogachev} 
	as well as the thermal neutron induced fission~\cite{Rochman1973}. The work shows the validity of isospin conservation in the neutron-rich
	 heavy nuclei~\cite{tifr,akjainPRL} and also confirms the theoretical predictions of Lane and Soper. This finding may lead to a new and better understanding of phenomena involving neutron-rich nuclei.

       %%%%%%%%%%%%%%%%%%%%%%%%%%%%%%%%%%%%%%%%%%%%%%%%%%%%%%%%%%%%%%%%%%%%%%%%%%%%%%%%%%%%%%%%%%%%%%%%%%%%%%%%%%%%%%%%%%%%%%%%%%%%%%%%%%%%%%%%%%%%%%%%%%%%%%%%%%%

       \section{METHODOLOGY}

       We adopt the convention that the nucleons carry an isospin $t=1/2$ with projections $t_3=+1/2$ (for neutron) and $t_3=-1/2$ (for proton). 
	The projection of the total isospin of the nucleus is given by $T_3=\sum t_3=(N-Z)/2$~\cite{Casten}. The total isospin $T$ can have any 
	value between $(N-Z)/2$ to $(A/2)$.
       
       We consider an induced fusion-fission reaction where the compound nucleus (CN) formed by the fusion of the projectile (P) and the target (T) fissions into two fragments $F_1$ and $F_2$ with the emission of q neutrons,
       \begin{eqnarray}
       P(Z_P,N_P) &+& T(Z_T,N_T) \rightarrow CN(Z,N)^* \nonumber \\ 
       %\downarrow \nonumber \\
       &\rightarrow& F_1(Z_1,N_1)+F_2(Z_2,N_2)+q~n .
       \label{CN}
       \end{eqnarray}
       Conservation of $T_3$ implies that
       \begin{eqnarray}
       T_{3_P} + T_{3_T} = T_{3_{CN}} = T_{3_{F_1}} + T_{3_{F_2}} + q/2 ~.
       \label{N-Z}
       \end{eqnarray}
       The total isospin of the projectile $T_P$, and the target $T_T$, are considered to have their minimum value, i.e. $T_P=T_{3_P}$, $T_T=T_{3_T}$. Hence, the isospin of the CN lies in the range $\left|{T_T-T_P}\right| \le T_{CN} \le \left({T_T+T_P} \right)$. Since $T_{3_{CN}}=T_{3_T}+T_{3_P}=T_T+T_P$, the $T_{CN}$ can have only one unique value, i.e., $T_{CN}=T_T+T_P$. Since $T_{CN}$ has a unique value and the emitted neutrons have pure isospin ($=q/2$), the total isospin of the fission fragments must lie in the range 
       \begin{eqnarray}
       \left|{T_{CN}-{(q/2)}}\right| \le T_{F_1}+T_{F_2} \le \left({T_{CN}+{(q/2)}} \right) .
       \label{CN-1}
       \end{eqnarray}
       To facilitate the discussion, we introduce an auxiliary concept of a residual compound nucleus (RCN), formed after the emission of the neutrons from the CN with $T_{RCN}=T_{F_1}+T_{F_2}$. Further, for a given number of emitted neutrons, the isospin components $T_3$ of the correlated pairs of fragments in a given $Z_1,Z_2$ partition are required to form a multiplet of the total isospin $T$ of the fragments.
       In a single partition, the isospin part of the RCN wave function is related to the isospins of the two correlated fission fragments as
       \begin{eqnarray}
       |T,T_3\rangle_{RCN} = \sum \langle CGC \rangle |T_{F_1},T_{3_{F_1}}\rangle|T_{F_2},T_{3_{F_2}}\rangle ,
       \label{CGCoeff0}
       \end{eqnarray}
       where the sum extends over all the allowed possibilities. Here, $\langle CGC \rangle$ denotes the Clesbch Gordon Coefficients 
	given by $\langle T_{F_1}T_{F_2}T_{3_{F_1}}T_{3_{F_2}} | T_{RCN}T_{3_{RCN}} \rangle$. The relative yield or intensity of each member of the correlated fragment pairs within a given partition can be obtained from the square of the corresponding $\langle CGC \rangle$.

       %%%%%%%%%%%%%%%%%%%%%%%%%%%%%%%%%%%%%%%%%%%%%%%%%%%%%%%%%%%%%%%%%%%%%%%%%%%%%%%%%%%%%%%%%%%%%%%%%%%%%%%%%%%%%%%%%%%%%%%%%%%%%%%%%%%%%%%%%%%%%%%%%%%%%%%%%

       \section{RESULTS AND DISCUSSIONS}
       
       In the present paper, we show the results of our study on two recent measurements of the HI fusion-fission reactions, 
	namely $^{238}$U($^{18}$O,f) ~and $^{208}$Pb($^{18}$O,f)~\cite{Danu,Bogachev} and a thermal neutron induced fission 
	reaction $^{245}$Cm(n$^{th}$,f).
       
       %%%%%%%%%%%%%%%%%%%%%%%%%%%%%%%%%%%%%%%%%%%%%%%%%%%%%%%%%%%%%%%%%%%%%%%%%%%%%%%%%%%%%%%%%%%%%%%%%%%
       
       We first consider the reaction $^{238}$U($^{18}$O,f).
       Here, $T_{3_{T}}=27$, $T_{3_{P}}=1$. From the previous arguments, the isospin of the CN , $^{256}_{100}$Fm, is uniquely fixed as $T_{CN}=T_T+1=28$.  
       The CN decays into seven distinct partitions namely Sn-Sn, Cd-Te, Pd-Xe, Ru-Ba, Mo-Ce, Zr-Nd and Sr-Sm correlated pairs of fragments~\cite{Danu}. 
	The experimental study shows that the $12$ and $8$ neutron emission channels have the maximum yield in the case of the Sn-Sn and Ru-Ba 
	partitions respectively. For all the other partitions, the $10n$ emission channel has the maximum yield. Here, we focus only on these dominant channels.
       
\begin{figure}[h]
       \centering
       \includegraphics[scale=0.55,trim=0cm 0cm 0cm 0cm,clip]{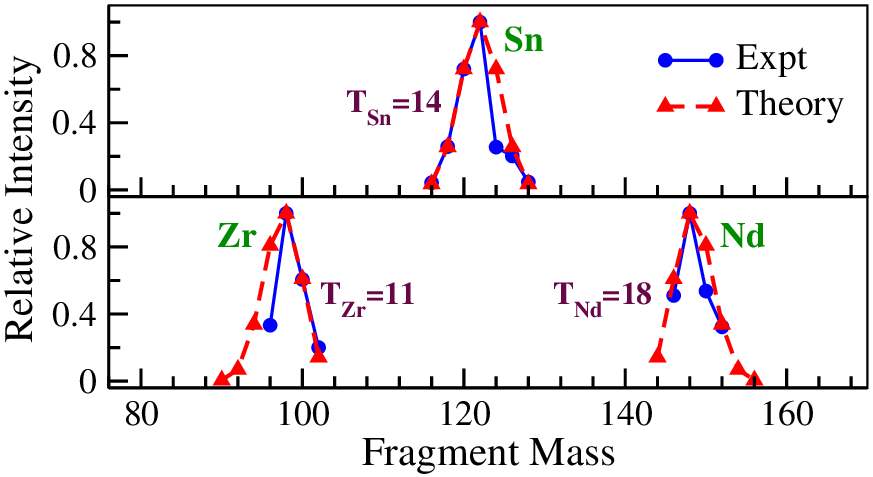}
       \caption{Calculated and measured\cite{Danu} relative intensities of the correlated pairs of fission fragments in two partitions for the reaction $^{238}$U($^{18}$O,f).}
       \label{fig:fig2}
       \end{figure}

       For the partitions with 8, 10, and 12 neutron emission channels dominating the yield, $T_{3_{RCN}}~=~28-4=24,~28-5=23,~28-6=22$ respectively. 
	For those partitions where $8$ or $12$ neutron emission channels have the maximum yield, i.e. the 8 or 12 evaporating neutrons carry away 
	even values of $T_3$ and $T$, so that the RCN has an even value of $T_3$, we assign $T_{RCN}=28$ (an even value) which is the minimum value of 
	$T$ that can generate all the observed correlated fragment pairs in a partition as members of an isospin multiplet. In those cases where $10$ neutron emission channel has the dominant yield leading to an odd value of $T_3$ for the RCN, we assign $T_{RCN}=29$, which is the minimum odd value of $T_{RCN}$ that can generate all the observed fragment pairs as members of isospin multiplets.
       For the Sn-Sn partition, the RCN (assigned $|T_{RCN},T_{3_{RCN}}\rangle = |28,22\rangle$) breaks up into correlated pairs of Sn fragments as:
       \begin{eqnarray}
       |28,22\rangle_{RCN} = \sum \langle CGC \rangle | T',T_3'\rangle_{Sn}|T'',T_3''\rangle_{Sn}
       \label{CGCoeff}
       \end{eqnarray}
       where, $\langle CGC \rangle = \langle 14,14,T_{3_{Sn}}' T_{3_{Sn}}'' | 28,22 \rangle$. The assignment of the $T$ values to the RCN and 
	the fragment pairs in each partition satisfy isospin conservation. The larger $T$ values used are supported by the arguments given by 
	Kelson~\cite{Kelson}. A remarkable agreement is obtained between the calculated and the observed relative intensities (both normalized 
	to $1$ at the peak value) of all the partition-wise correlated pairs of even-even fission fragments 
	(two of the cases shown in Fig.~\ref{fig:fig2}). This suggests that the $T$ values assigned as above appear to make the maximum 
	contribution to the relative yields, though other $T$ values in the allowed range are also possible. Sudden dips in the measured 
	intensities for few mass numbers of certain nuclei (eg. $^{124}$Sn) can be accounted for by the presence of isomeric states and/or structure effects.

       %%%%%%%%%%%%%%%%%%%%%%%%%%%%%%%%%%%%%%%%%%%%%%%%%%%%%%%%%%%%%%%%%%%%%%%%%%%%%%%%%%%%%%%%%%%%%%%%%%%%%%%%%%%55

       We, now, show the calculations for a couple of partitions from the reaction $^{208}$Pb($^{18}$O,f)~\cite{Bogachev}. The CN, $^{226}$Th, 
	formed by the fusion of the projectile $^{18}$O ($|1,1\rangle$), with the target $^{208}$Pb ($|22,22\rangle$), 
	has $|T_{CN},T_{3_{CN}}\rangle = |23,23\rangle$. Six different partitions consisting of Ru-Pd, Mo-Cd, Zr-Sn, Sr-Te, Kr-Xe and Se-Ba 
	correlated fragments were observed. Though the measured n-multiplicity distributions for the six partitions are wide, the $4-6$ neutron 
	emission channels have maximum yields~\cite{Bogachev} and are considered here. 
       
       \begin{figure}[h]
       \centering
       \includegraphics[scale=0.55,trim=0cm 0cm 0cm 0cm,clip]{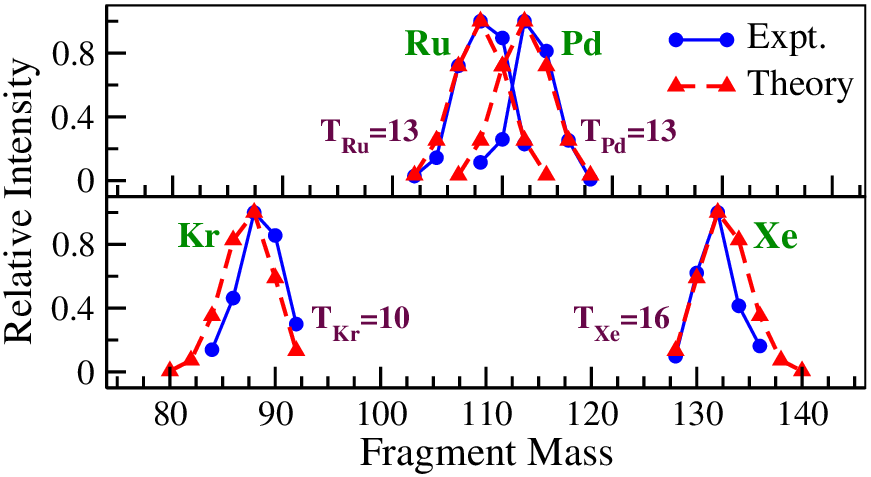}
       \caption{Calculated and measured ~\cite{Bogachev} relative intensities of the correlated pairs of fission fragments in two partitions for the reaction $^{208}$Pb($^{18}$O,f).}
       \label{fig:fig3}
       \end{figure}
       
       We again follow a similar line of arguments of assigning $T_{RCN}$ as earlier. We present the calculations and results for 
	a couple of partitions i.e. the most symmetric (Ru-Pd) and a highly asymmetric one (Kr-Xe). For both the partitions, the 6n emission channel dominates the yield assigning $|T_{RCN},T_{3_{RCN}}\rangle = |26,20\rangle$. Hence, we have
       \begin{eqnarray}
       |26,20\rangle_{RCN} = \sum \langle CGC ~\rangle | 13,T_3\rangle_{Ru}~|13,T_3\rangle_{Pd} \\
       |26,20\rangle_{RCN} = \sum \langle CGC \rangle ~| 10,T_3\rangle_{Kr}~|16,T_3\rangle_{Xe}
       \label{CGCoeff-3}
       \end{eqnarray}
       An excellent agreement is again obtained between the calculated and the measured intensities as shown in Fig.~\ref{fig:fig3}.

       %%%%%%%%%%%%%%%%%%%%%%%%%%%%%%%%%%%%%%%%%%%%%%%%%%%%%%%%%%%%%%%%%%%%%%%%%%%%%%%%%%%%%%%%%%%%%%%%%%%%%%%%%%%5

       We now discuss the thermal n-induced fission $^{245}$Cm(n$^{th},f)$~\cite{Rochman1973}.
       The CN, $^{246}$Cm, formed by the fusion of the neutron ($|\frac{1}{2},\frac{1}{2}\rangle$), with the target $^{245}$Cm 
	($|\frac{53}{2},\frac{53}{2}\rangle$), has $|T_{CN},T_{3_{CN}}\rangle = |27,27\rangle$.
       \begin{table}[h]
       %\caption{\label{tab:table1}$T$ the emitted neutrons ($T_{ns}$) and the RCN along with the weight factors for the various neutron emission channels.}
       \caption{\label{tab:table1}Weight factors for the various neutron emission channels along with the corresponding $T_{ns}$ (isospin carried away by the neutrons) and $T_{RCN}$.}
       %\begin{tabular*}{0.35\textwidth}{@{\extracolsep{\fill}}|cccc|}
       \begin{tabular}{|c|c|c|c|}
       \hline 
       neutron emission & weight factor & $T_{ns}$ & $T_{RCN}$ \\ \hline
       %emission & factor & &  \\ \hline
       %& & neutrons & \\ \hline
       0 & 0.004373 & 0 & 27 \\
       1 & 0.008011 & 0.5 & 27.5 \\
       2 & 0.100233 & 1 & 27  \\
       3 & 0.277928 & 1.5 & 28.5 \\
       4 & 0.334261 & 2 & 28  \\
       5 & 0.196610 & 2.5 & 29.5  \\
       6 & 0.065010 & 3 & 29  \\
       7 & 0.017510 & 3.5 & 28.5  \\
       \hline
       \end{tabular}
       \end{table}
       The authors~\cite{Rochman1973} have measured the isotopic yields from mass $A=85$ to $115$ with $Z=33$ to $47$. The corresponding heavier members of the fission fragments for these nuclei could not be observed due to the experimental limitation. Here, we discuss the relative yield of the 
	correlated pairs of fragments for two of the fission partitions, i.e., Ru-Te (only Ru isotopes observed) and Kr-Nd (only Kr isotopes observed).
       \begin{figure}
       \includegraphics[scale=0.55,trim=0cm 0cm 0cm 0cm,clip]{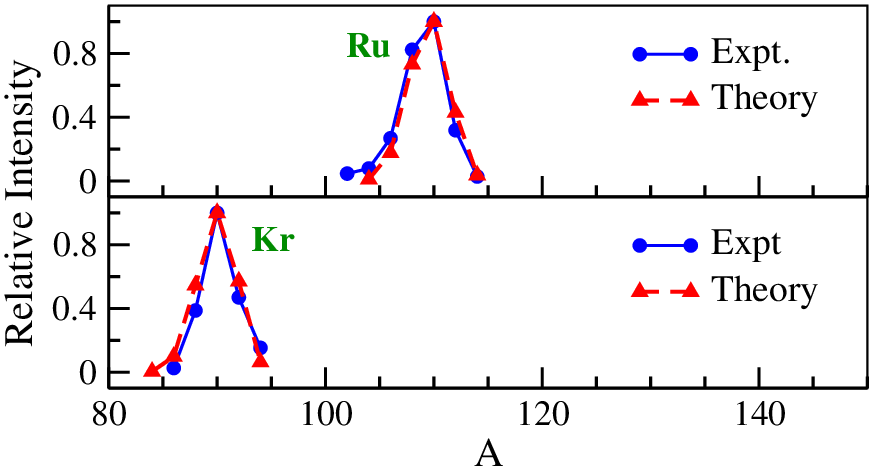}% Here is how to import EPS figure if you are using 
       \caption{\label{fig1} Calculated and measured~\cite{Rochman1973} relative yields of the observed even-even isotopes of Ru (top) and Kr (bottom) obtained from the reaction $^{245}$Cm(n$^{th}$,f).}
       \end{figure}
       In thermal neutron induced reactions, $0-7$ neutrons are emitted with the respective weight factors which we have considered in our 
	interpretation/calculations (see Table~\ref{tab:table1}). The weight factors used have been taken from the LLNL report~\cite{LLNL} and 
	correspond to the same number of average neutrons emitted. The $T$ values assigned to the fission fragment pair depend on the $T$ value taken 
	away by the emitted neutrons as mentioned before. An excellent agreement between the calculated and experimental results (see Fig.~\ref{fig1}) 
	is again observed. This verifies that the concept of isospin conservation is equally valid and important in the neutron-induced fission reaction as it is in the HI induced fusion-fission reactions. Further, for all the three cases discussed above, we note that, for a particular neutron-emission channel, the more asymmetric correlated pairs of fission fragments are related to more asymmetric combinations of the individual $T$ values.

       %%%%%%%%%%%%%%%%%%%%%%%%%%%%%%%%%%%%%%%%%%%%%%%%%%%%%%%%%%%%%%%%%%%%%%%%%%%%%%%%%%%%%%%%%%%%%%%%%%%%%%%%%%%%%%%%%%%%%%%%%%%%%%%%%%%%%%%%%%%%%%%%%%%%%%%%%%%%

       \section{SUMMARY}
       
       Three induced fission reactions have been considered and the relative yield of the correlated pairs of the n-rich fission fragments in each of the partitions have been calculated using isospin conservation. It has been found that the isospin conservation can precisely explain the relative yields of the fission fragment pairs in each partition. The findings give the first direct and firm evidence of isospin conservation in neutron rich nuclei and the reactions involving such nuclei.

       %%%%%%%%%%%%%%%%%%%%%%%%%%%%%%%%%%%%%%%%%%%%%%%%%%%%%%%%%%%%%%%%%%%%%%%%%%%%%%%%%%%%%%%%%%%%%%%%%%%%%%%%%%%%%%%%%%%%%%%%%%%%%%%%%%%%%%%%%%%%%%%%%%%%%%%%%%%%

       \section{ACKNOWLEDGEMENTS}
       
       The authors thank Profs. Don Robson, P. Van Isacker and V. K. B. Kota for several useful discussions. Financial support from the D.S.T, D.A.E. and M.H.R.D. (Govt. of India) is acknowledged.

       %%%%%%%%%%%%%%%%%%%%%%%%%%%%%%%%%%%%%%%%%%%%%%%%%%%%%%%%%%%%%%%%%%%%%%%%%%%%%%%%%%%%%%%%%%%%%%%%%%%%%%%%%%%%%%%%%%%%%%%%%%%%%%%%%%%%%%%%%%%%%%%%%%%%%%%%%%%%

       \end{document}